\documentclass[aps,prl,twocolumn,showpacs,groupedaddress]{revtex4}  
\usepackage{graphicx}  
\usepackage{dcolumn}   
\usepackage{bm}        
\usepackage{amssymb}   

\def\d0        {D\O}
\def\pp        {p\bar{p}}
\def\bb        {b\bar{b}}
\def\ttop      {t\bar{t}}
\def\tt        {\tau^+\tau^-}

\def\tanb      {\tan\!\beta}
\def\tanbsq    {\tan^2\!\beta}
\def\mAtanb    {(m_A,\tan\!\beta)}
\newcommand{\met}{{\ooalign{\hfill$E_T\!\!$\hfill\crcr\hspace{1pt}\large{/}}}~}

\begin{document}


\hspace{5.2in} \mbox{FERMILAB-PUB-08/451-E}
\title{Search for neutral Higgs bosons at high {\boldmath$\tanb$}       
       in the {\boldmath$b(h/H/A)\to b\tt$} channel}
%
\author{V.M.~Abazov$^{36}$}
\author{B.~Abbott$^{75}$}
\author{M.~Abolins$^{65}$}
\author{B.S.~Acharya$^{29}$}
\author{M.~Adams$^{51}$}
\author{T.~Adams$^{49}$}
\author{E.~Aguilo$^{6}$}
\author{M.~Ahsan$^{59}$}
\author{G.D.~Alexeev$^{36}$}
\author{G.~Alkhazov$^{40}$}
% ...

\affiliation{\vspace{0.1 in}(The D\O\ Collaboration)\vspace{0.1 in}}
% ...
\affiliation{$^{6}$University of Alberta, Edmonton, Alberta, Canada,
                Simon Fraser University, Burnaby, British Columbia, Canada,
                York University, Toronto, Ontario, Canada, and
                McGill University, Montreal, Quebec, Canada}
% ...
\affiliation{$^{29}$Tata Institute of Fundamental Research, Mumbai, India}
% ...
\affiliation{$^{36}$Joint Institute for Nuclear Research, Dubna, Russia}
% ...
\affiliation{$^{40}$Petersburg Nuclear Physics Institute,
                St. Petersburg, Russia}
% ...
\affiliation{$^{49}$Florida State University, Tallahassee, Florida 32306, USA}
% ...
\affiliation{$^{51}$University of Illinois at Chicago,
                Chicago, Illinois 60607, USA}
% ...
\affiliation{$^{59}$Kansas State University, Manhattan, Kansas 66506, USA}
% ...
\affiliation{$^{65}$Michigan State University,
                East Lansing, Michigan 48824, USA}
% ...
\affiliation{$^{75}$University of Oklahoma, Norman, Oklahoma 73019, USA}
% ...
\date{October 31, 2008}
           
\begin{abstract}
The first search in $\pp$ collisions at $\sqrt{s}=1.96$ TeV
for the production of neutral Higgs bosons 
in association with bottom quarks and decaying in two tau leptons
is presented. 
The cross section for this process is enhanced in many extensions of the standard model
(SM), such as its minimal supersymmetric extension (MSSM) at large $\tanb$.
The data, corresponding to an integrated luminosity of 328 pb$^{-1}$, were collected
with the D0 detector at the Fermilab Tevatron Collider.
An upper limit is set on the production cross section of neutral Higgs
bosons in the mass range of 90 to 150 GeV,
and this limit is used to exclude part of the MSSM parameter space.
\end{abstract}

\maketitle

In the minimal supersymmetric extension of the standard model (MSSM),
the Higgs sector consists of five physical Higgs bosons:
two neutral scalars, $h$ and $H$ (with $m_h < m_H$ by convention),
one neutral pseudo-scalar, $A$,
and a charged pair, $H^{\pm}$.
At leading order (LO), the coupling of the neutral Higgs bosons to down-type quarks
is proportional to $\tanb$,
where $\tanb$ is the ratio of the vacuum expectation values of the two Higgs doublets.
The production cross section of a neutral Higgs boson in association
with a down-type quark, such as the $b$ quark, is therefore proportional to $\tanbsq$ (at LO).
Thus, the $b\phi$ ($\phi=h, H, A)$ production mechanism provides a natural mode
to search for a neutral Higgs boson at high $\tanb$ in the MSSM~\cite{CHWW}.

In most of the MSSM parameter phase space, the neutral scalar Higgs bosons $h$ and $H$
decay $\sim$90\% of the time into a pair of $b$ quarks,
and $\sim$10\% of the time into a pair of tau leptons.
The neutral pseudoscalar $A$ decays into $\bb$ or $\tt$ in all of the parameter space,
with similar branching ratios ($\sim$90\% and $\sim$10\%, respectively). 
In this Letter, we present a search for the production of a neutral Higgs boson
in association with a $b$ quark,
with the subsequent decay of the Higgs boson into two tau leptons,
using data collected by the D0 experiment in $\pp$ collisions at $\sqrt{s}=1.96$ TeV
at the Fermilab Tevatron collider.
We perform the analysis using the final state where one tau decays leptonically
into a muon ($\tau \rightarrow \mu \nu_{\mu} \nu_{\tau}$),
and the other tau decays hadronically into a narrow jet 
($\tau \rightarrow \tau_h \nu_{\tau}$, where $\tau_h$ denotes the hadronic tau jet).

The $b\phi \rightarrow b\tt$ search channel is complementary to
the $b\phi \rightarrow bb\bar{b}$~\cite{bbb} and
the inclusive $\phi \rightarrow \tt$~\cite{TEVtautau} searches.
The $\tt$ decay mode of the Higgs boson is less sensitive than the $\bb$ decay
to the large supersymmetric radiative corrections on the production cross section and
decay width~\cite{CHWW}.
Experimentally, the $b\tt$ channel presents a clean signature which does not suffer
from the large heavy-flavor multi-jet background of the $bb\bar{b}$ channel,
and is less affected by the $Z \rightarrow \tt$ background than the inclusive
$\phi \rightarrow \tt$ channel.

The D0 detector~\cite{D0detector} consists of
a central tracking system, comprising a silicon microstrip tracker and a central 
fiber tracker, both within a 2 T solenoidal magnet;
a liquid-argon and uranium calorimeter, divided into a central calorimeter
and two end calorimeters;
and a muon system, consisting of three layers of tracking detectors and
scintillation trigger counters.

This analysis considers data collected by the D0 experiment between August 2002 and June 2004.
Two single-muon triggers are used, requiring a muon with transverse momentum ($p_T$)
greater than either 3 or 5 GeV and a track with $p_T>10$ GeV.
The total integrated luminosity for the selected triggers is $328\pm20$ pb$^{-1}$~\cite{NewLumi}.

Signal events are simulated using the process
$\pp \rightarrow b\phi \rightarrow b\tt$ in {\sc pythia}~\cite{pythia}, 
where one of the tau leptons is forced to decay leptonically into a muon
and the second tau is free to decay to all allowed modes;
the $b$ quark is generated with $p_T>15$ GeV and $|\eta|<2.5$, where 
$\eta = -\ln[\tan(\theta/2)]$ 
is the pseudorapidity and $\theta$ is the polar angle relative to the proton beam 
direction. 
Background processes such as $t\bar{t}$, $W$+jets and $WW$ production
are simulated using {\sc alpgen}~\cite{alpgen} interfaced with {\sc pythia}
for showering and fragmentation. Additional $p\bar{p}$ interactions are modeled
with {\sc pythia} according to a Poisson distribution with mean of 0.4 events,
which corresponds to the expected average multiplicity in the data.
The simulated events are processed through a GEANT-based~\cite{geant} simulation 
of the D0 detector and reconstructed with the same software as the collider data.
They are also weighted on an event-by-event basis by the trigger efficiency 
parametrization measured in the data.
The trigger efficiency, estimated on the simulated signal sample
after selecting $\mu\tau_h$ pairs, is $(62\pm1)$\%.

There are three types of physics objects used in this analysis: muons,
hadronic taus, and jets. All selected objects are required to be associated
with the same primary vertex within 1 cm along the beam direction.

Muons are reconstructed from patterns of hits in the muon detectors matched to isolated
central tracks, and are required to have $p_T>12$ GeV.

Hadronically decaying taus are characterized by a narrow isolated jet with low track
multiplicity. We distinguish three tau types:
(1) a single track with energy deposited in the hadronic calorimeter,
(2) a single track with energy deposited both in the hadronic and electromagnetic calorimeters, and
(3) three tracks with corresponding energy deposited in the calorimeter.

After an initial selection of tau candidates based on the transverse energy ($E_T$)
of the calorimeter cluster,
sum of the track transverse momenta, and isolation and width of the associated 
calorimeter energy deposits, the candidates are
further discriminated against jets using a neural network (NN) which has been trained 
separately for each tau type~\cite{Z-tautau}.  
For types 1 and 2, tau candidates are required to have a NN output greater
than 0.8.  For type 3 tau candidates, because of the larger multijet background, 
the NN selection is tightened to 0.98. 
The average tau identification efficiency in signal events is $\sim$62\%. 

Jets are reconstructed from clusters of energy in the calorimeter using 
the D0 Run II midpoint cone algorithm with a radius of 0.5~\cite{jetalg}.
Jet energies are corrected to the particle level.
Events are required to have at least one jet identified as originating from
a $b$ quark ($b$ tagged) and with $p_T>15$ GeV and $|\eta|<2.5$.
Jets are $b$ tagged using an algorithm that computes
the probability that the jet originated from a $b$ hadron,
based on the impact parameter of the tracks associated with the jet~\cite{JLIP}.
For a jet of $p_T=20$ GeV and $|\eta|<2.5$, as is typical for signal events,
the $b$-tagging efficiency measured in data is $\sim$40\%,
whereas the probability to tag a light-flavor jet is $\sim$1\%.
A parameterization of the $b$-tagging efficiency measured in data
is applied to each simulated jet, according to its $p_T$, $\eta$ and flavor.

Main backgrounds to the $b\phi \rightarrow b\tt \rightarrow b\mu\tau_h$ 
process are multijet, $Z$+jets and $t\bar{t}$ production. 
Smaller background contributions originate from $W$+jets and $WW$ production.
The multijet and $Z$+jets backgrounds are estimated from the data, whereas
all other backgrounds are estimated from the Monte Carlo (MC) simulation.

A multijet background event typically consists of two or more jets, with one
jet misidentified as a hadronic tau, a real or misidentified $b$ jet,
and a muon from a heavy-flavor decay that appears isolated.
Since the charge of the muon is not correlated with the charge of the hadronic tau candidate,
the multijet background tends to have equal amounts of opposite-sign (OS) 
and same-sign (SS) $\mu\tau_h$ pairs.
In contrast, the signal should contain only opposite-sign $\mu\tau_h$ pairs coming 
from the Higgs decay.  
Thus, we require that the reconstructed muon and hadronic tau have opposite charges.
The multijet background in the OS sample is estimated from the SS events in the data as follows: 
first, the SS yield is corrected for non-multijet backgrounds by subtracting these 
based on MC estimates;
second, the corrected SS multijet yield is multiplied by the probability of
a jet to be misidentified as a hadronic tau; 
third, a correction is applied to account for a small
asymmetry observed in OS and SS multijet control samples;
finally, the probability of a multijet event to have at least
one $b$-tagged jet is applied.

The production of a $Z$ boson in association with jets contributes as
a background via $Z \rightarrow \tau^+\tau^- \rightarrow \mu\tau_h$ and
$Z \rightarrow \mu^+\mu^-$ decays, and where one of the jets is 
a real or misidentified $b$ jet. In the case of $Z \rightarrow \mu^+\mu^-$,
one of the muons is misidentified as a hadronic tau.
The contribution from both real and misidentified $b$ jet backgrounds,
in either $Z$ decay channel, is estimated by measuring the fraction
of $b$-tagged events in $Z \rightarrow \mu^+\mu^-$ data, found to be
$(2.5\pm0.4)$\%, and multiplying it by the estimated number of 
$Z(\rightarrow\mu\tau_{h})$+jets events in data before $b$ tagging.

After $b$ tagging, $t\bar{t}$ production is the dominant background.
Such events are characterized by having higher $p_T$ objects than those
in signal events. Therefore, in order to reduce the $t\bar{t}$ background,
we use a neural network (KNN) which exploits kinematic differences between
signal and background, based on four variables:
the sum of the transverse momenta of all jets in the event (excluding the tau jet),
the missing transverse energy $\met$ (constructed from calorimeter
cells and the momenta of muons, and corrected for the energy response of taus and jets),
the jet multiplicity, and the azimuthal angular separation between the muon and the tau jet.
The neural network training is performed using a background MC sample of 
$t\bar{t}$ events where both $W$ bosons decay leptonically ($t\bar{t}\to\mu\tau_h$)
and a signal MC sample consisting of $b\phi \rightarrow b\tau^+\tau^- \rightarrow b\mu\tau_h$ 
events with a mixture of different Higgs masses.
In both samples, the events used passed all selection criteria except $b$ tagging. 
The KNN selection is optimized separately for each tau type. 
Events with type 1 and 3 taus have low $\ttop$ background and do not benefit from a
KNN selection. 
Requiring a KNN output greater than 0.4 has a signal efficiency of $\sim$95\% and
is found to be optimal for events with type 2 taus.
The amount of $t\bar{t}$ background remaining after the KNN selection is estimated from MC.

Systematic uncertainties affecting both signal and background predictions
based on MC are: integrated luminosity (signal: 6\%, background: $<$1\%)~\cite{NewLumi};
trigger efficiency (1.1\%);
tau identification (signal: 3-9\%, background: $<$0.4\%);
tau energy scale (10\%);
jet identification (signal: 6-9\%, background: $<$7\%);
jet energy scale (signal: 7-10\%, background: $<$4\%);
$b$-jet identification (signal: 5\%, background: $<$2\%);
and uncertainties on the signal (10\%) and $\ttop$, 
$W$+jets (20-30\%) and $WW$ theoretical cross sections. 
For backgrounds derived from data, the systematic uncertainties result 
from the limited statistics of the control data samples.  

\begin{figure*}[ht]
\begin{center}
\includegraphics[scale=0.29]{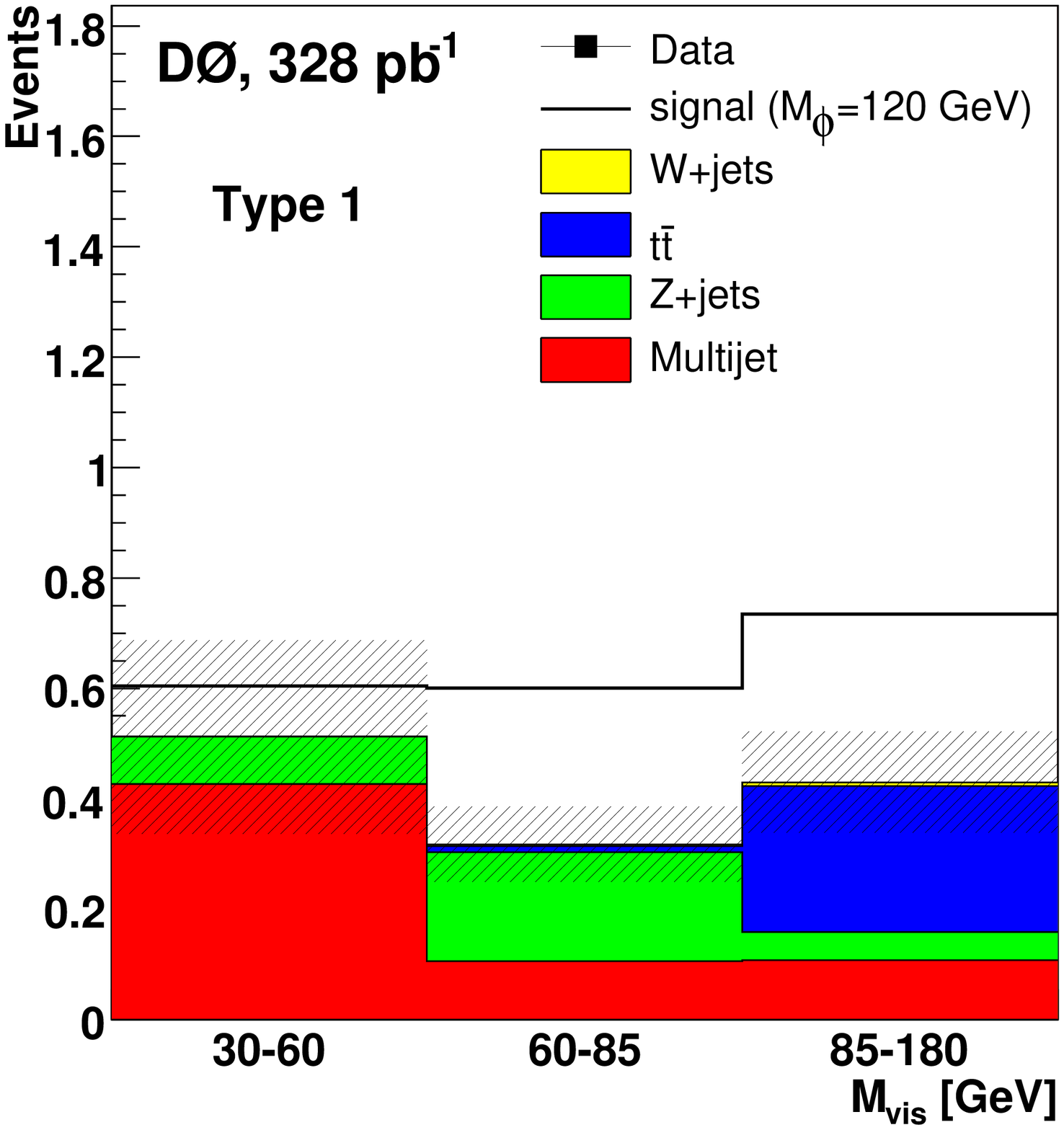}
\includegraphics[scale=0.29]{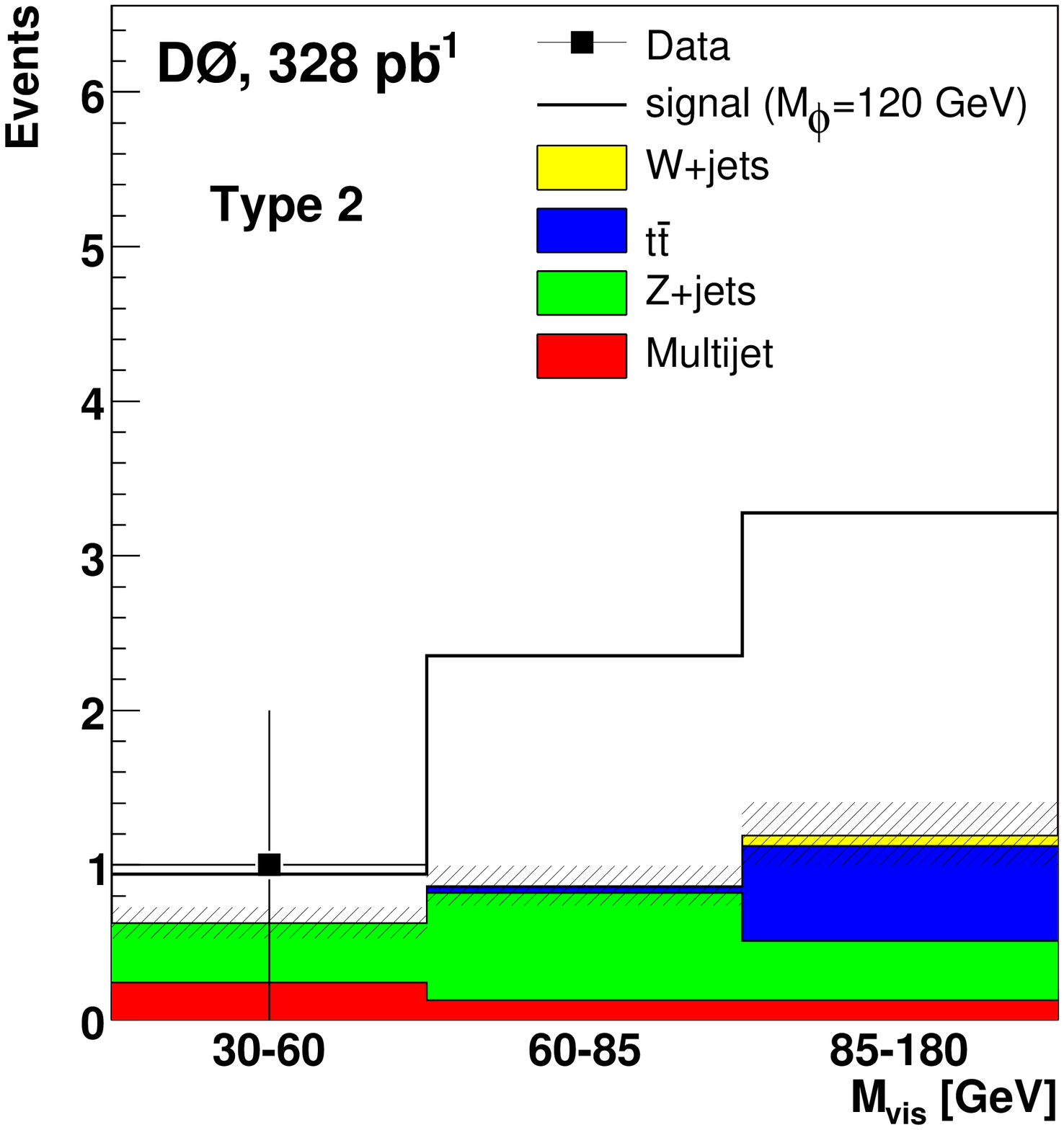}
\includegraphics[scale=0.29]{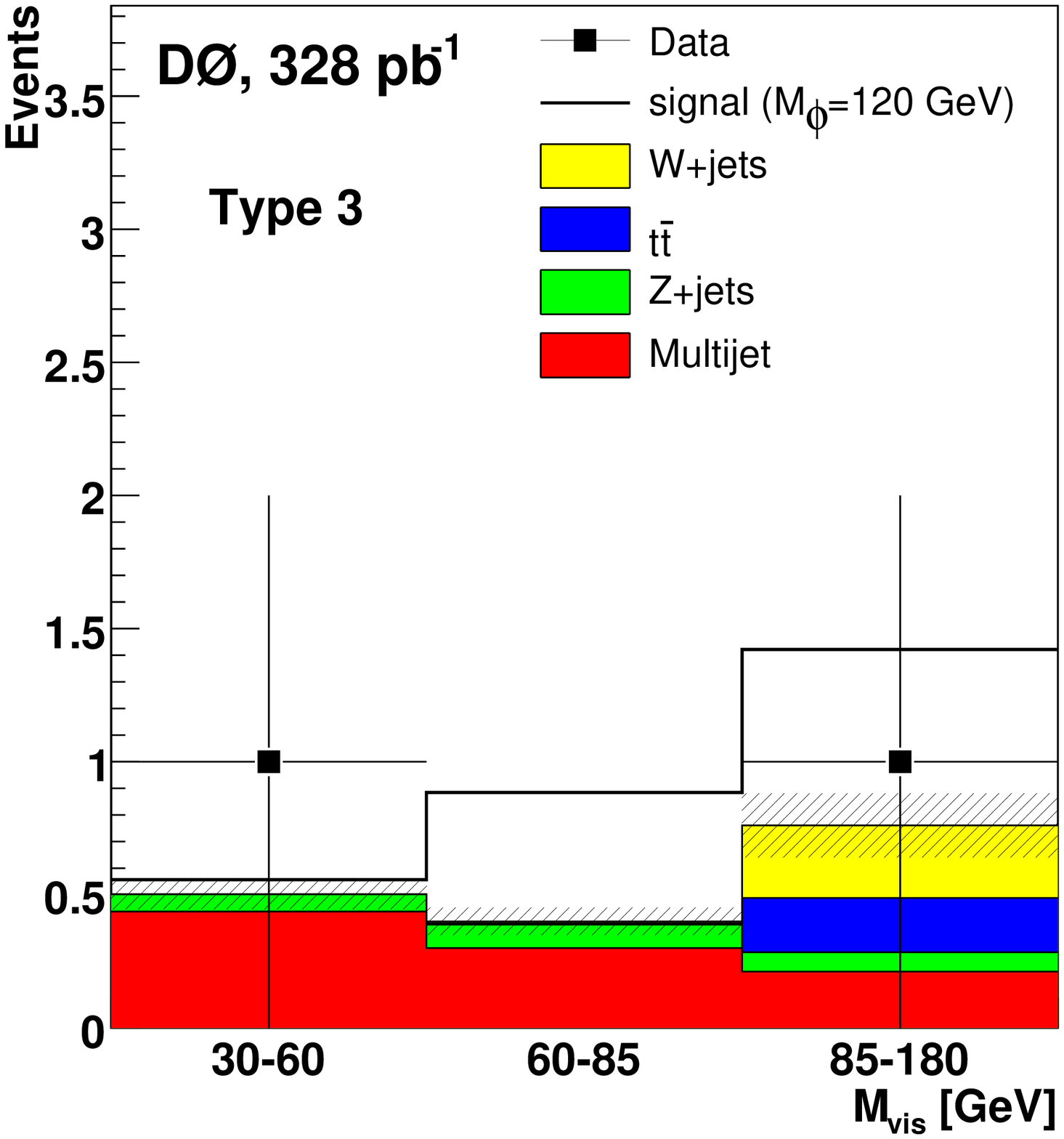}
\caption{Visible mass distributions for each tau type.
         Histograms show the signal and various backgrounds, points show the data.
         The error bands indicate the total uncertainty on the
         background estimation.}
\label{fig:mass}
\end{center}
\end{figure*}

The estimated number of events from the various backgrounds
and the observed number of events in the data
for the three tau types are presented in Table \ref{tab:sig_bkgd_obs}.
Also shown are the signal acceptance and the number of expected signal events 
for a Higgs mass $M_\phi=120$ GeV and $\tanb=80$.
The visible mass $M_{\rm vis}$ distributions, constructed from the four-vector momenta 
of the muon, hadronic tau, and missing momentum~\cite{TEVtautau},
for the data and SM prediction are shown in Fig.~\ref{fig:mass}.
No visible excess over the SM prediction is observed in the data.

\begin{table}[htbp]
\caption{\label{tab:sig_bkgd_obs}
         Expected number of events for backgrounds, number of observed events in
         data, signal acceptance for events with at least one muon and
         expected number of signal events for $M_\phi=120$ GeV and $\tanb=80$, for
         each hadronic tau type.
         Quoted uncertainties represent statistical and systematic
         added in quadrature.}
\begin{ruledtabular}
\begin{tabular}{cccc}
                     &      Type~1     &      Type~2     &      Type~3       \\ \hline
Multijet             & $0.60\pm0.22$   & $0.48\pm0.14$   & $0.95\pm0.16$     \\
$Z$+jets             & $0.34\pm0.09$   & $1.50\pm0.27$   & $0.25\pm0.08$     \\
$\ttop$              & $0.28\pm0.06$   & $0.65\pm0.18$   & $0.21\pm0.05$     \\
$W$+jets             & $0.009\pm0.005$ & $0.073\pm0.036$ & $0.28\pm0.12$    \\
$WW$                 & $0$             & $0.014\pm0.004$ & $0$               \\
Total Background     & $1.22\pm0.19$   & $2.71\pm0.33$   & $1.68\pm0.15$     \\ \hline
Observed             & $0$             & $1$             & $2$               \\ \hline
Signal Accept. (\%)  & $0.15\pm0.03$   & $0.87\pm0.14$   & $0.27\pm0.05$     \\   
Expected Signal      & $0.68\pm0.15$   & $3.9\pm0.7$     & $1.2\pm0.2$       \\ 
\end{tabular}
\end{ruledtabular}
\end{table}

\begin{figure*}[ht]
\includegraphics[scale=0.29]{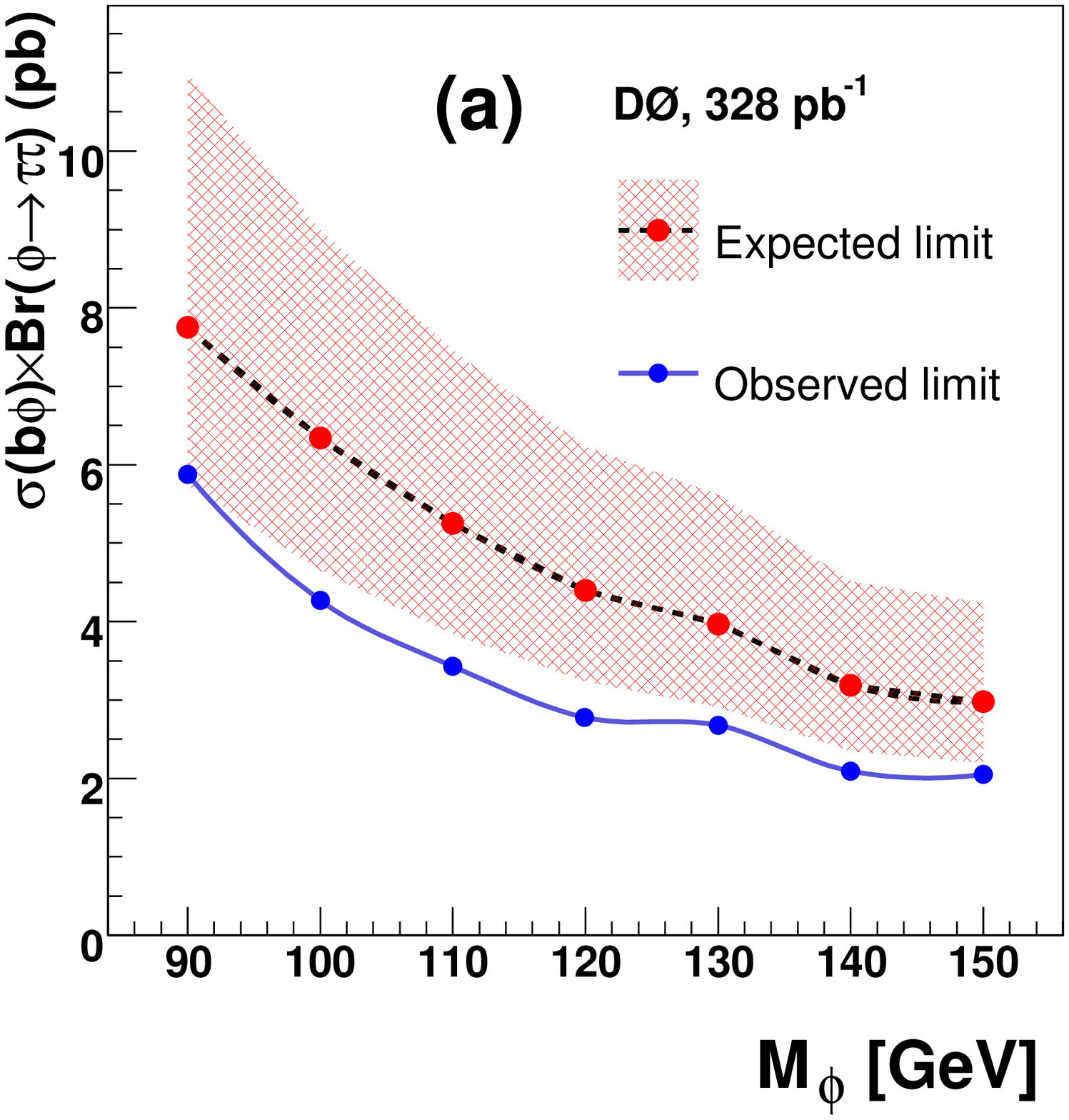}
\includegraphics[scale=0.29]{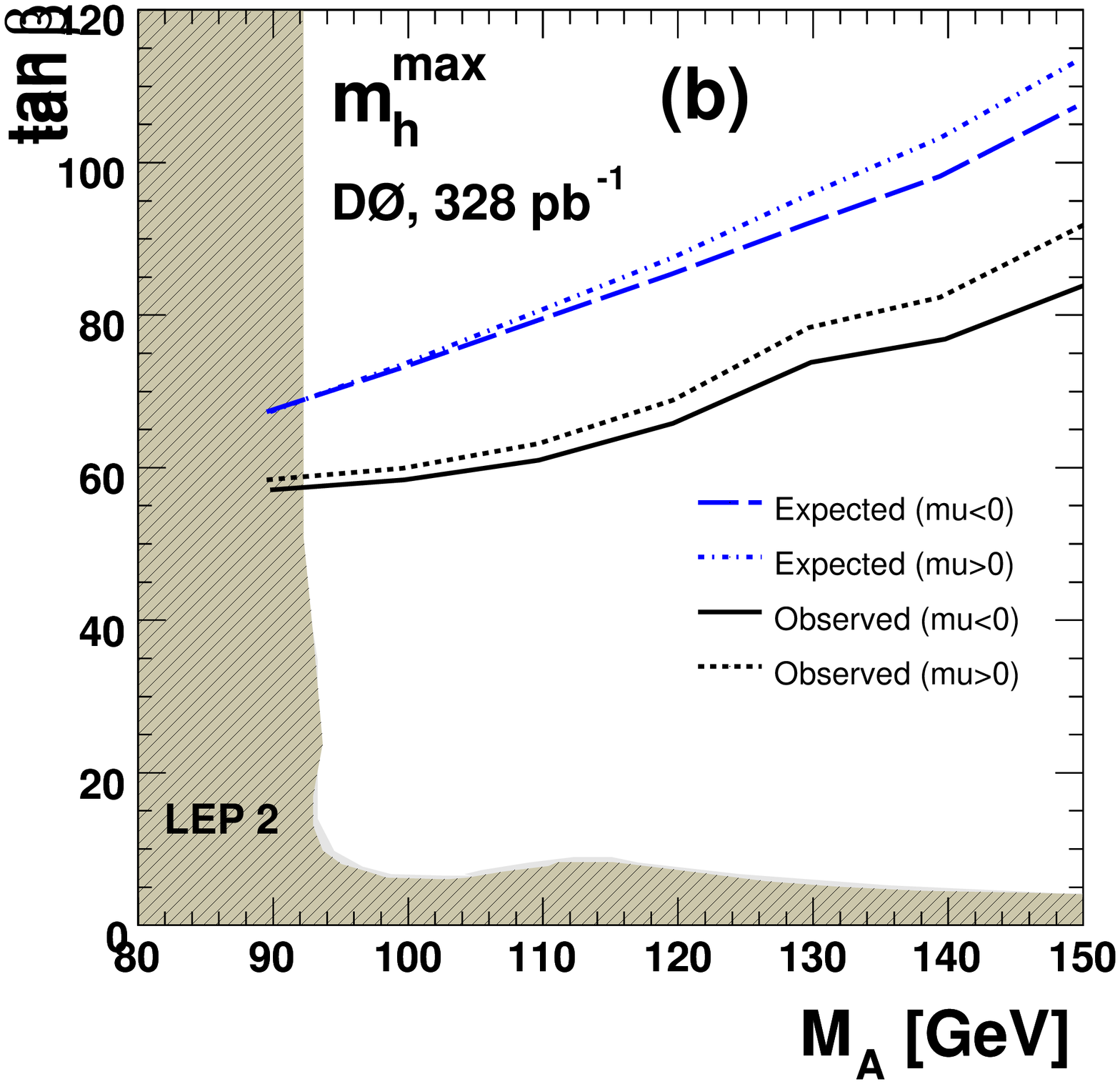}
\includegraphics[scale=0.29]{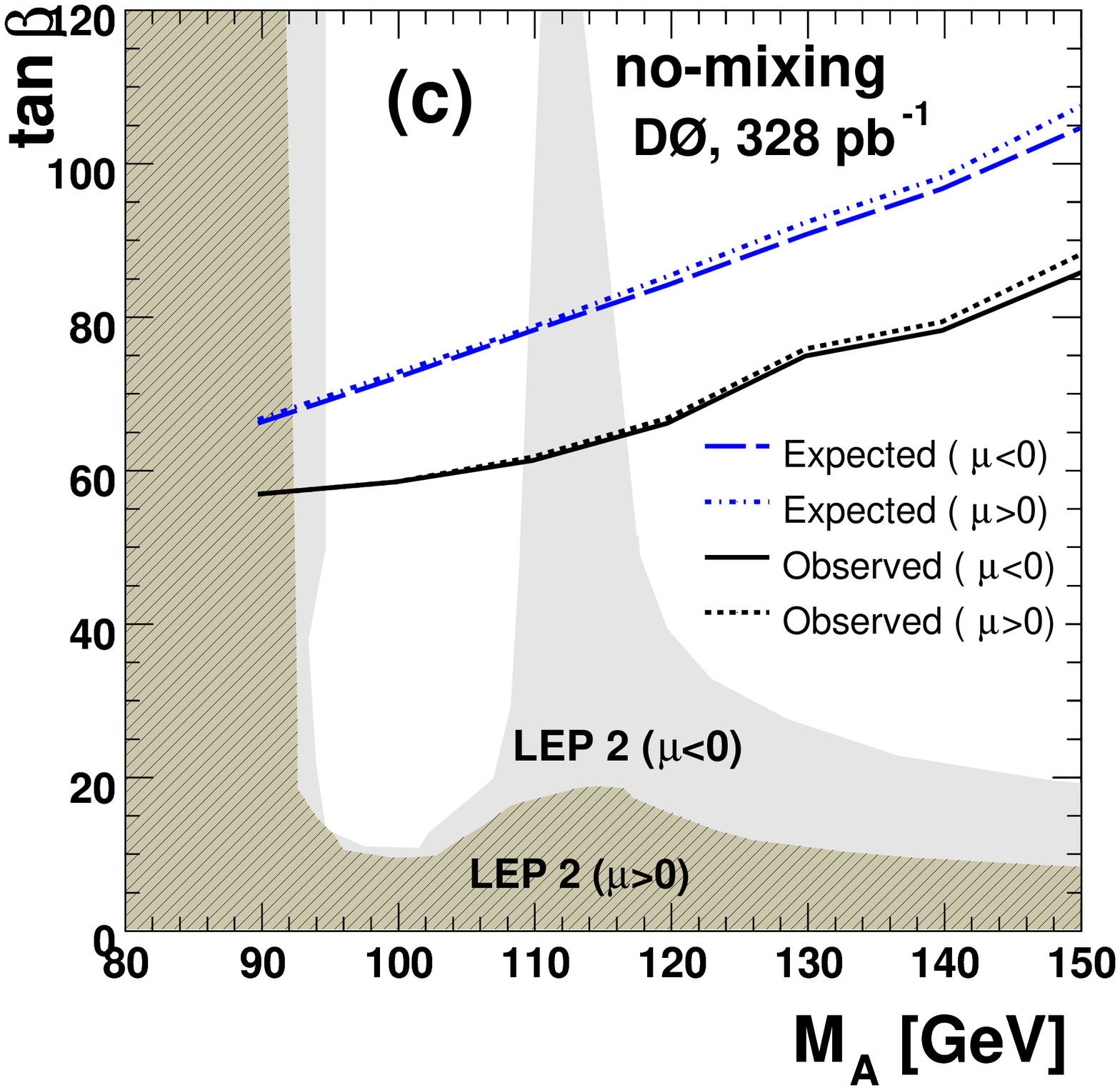}
\caption{
         (a) The 95\% C.L. expected and observed limits
         on the cross section times branching ratio
         for $\pp \rightarrow b\phi \rightarrow b\tau^+\tau^-$ production
         as a function of the Higgs mass. Also shown is the $\pm 1$ standard
         deviation band on the expected limit.
         These cross section limits are used to derive
         exclusion regions in the $\mAtanb$ plane
         for (b) the $m_h^{max}$ and (c) the no-mixing scenarios of the MSSM,
         for both $\mu = +200$ GeV and $\mu = -200$ GeV.
         Also shown is the region excluded by the LEP experiments.}
\label{fig:limits}
\end{figure*}

Upper limits on the production cross section times branching ratio
are set using a modified frequentist approach~\cite{Junk}.
In order to maximize the sensitivity, each tau type is treated as a separate channel
and the kinematic differences between signal and background are exploited by
using the $M_{\rm vis}$ distribution in the limit calculation.
In each channel, the $M_{\rm vis}$ distribution is split into three bins:
30-60, 60-85 and 85-180 GeV (see Fig.~\ref{fig:mass}).
The choice of bin size is driven by the available statistics in data 
to estimate the multijet background.
Figure~\ref{fig:limits} shows the 95\% confidence level (C.L.) upper limits on the production
cross section times branching ratio as a function of the Higgs mass.
Despite the $\sim$1:9 branching ratio of the $\tt$ to $\bb$ Higgs decay modes,
the upper limit on the $b\phi$ production cross section obtained by this analysis 
is competitive with the corresponding one in the $b\phi \rightarrow b\bb$ channel~\cite{bbb},
particularly at low $M_\phi$.

Using the cross section limit for $b\phi$ production, we can exclude regions of
$\mAtanb$ parameter space in the MSSM.
Beyond LO, the masses and couplings of the Higgs bosons in the MSSM depend
(through radiative corrections) on additional SUSY parameters, besides $m_A$ and $\tanb$.
Thus, we derive limits on $\tanb$ as a function of $m_A$ in two specific, commonly used
scenarios (assuming a CP-conserving Higgs sector):
the $m_h^{max}$ scenario and the no-mixing scenario~\cite{CHWW}.
The production cross sections, widths and branching ratios for the Higgs bosons
are calculated over the mass range 90-150 GeV using the {\sc mcfm} and {\sc feynhiggs}
programs~\cite{mcfm,feynhiggs}. 
Since at large $\tanb$ the $A$ boson is nearly degenerate in mass with either the $h$
or the $H$ boson, their production cross sections are added.
As shown in Fig.~\ref{fig:limits}, this analysis excludes a large portion 
of the MSSM parameter space.
For negative values of the \mbox{Higgsino} mass parameter $\mu$,
the $\tt$ decay mode explored here has comparable sensitivity to the $\bb$ decay mode~\cite{bbb}.
For positive values of $\mu$, however, the $\tt$ mode is superior to the $\bb$ mode,
as it does not suffer from the effect of the large supersymmetric radiative corrections 
to the Higgs production cross section and decay width~\cite{CHWW}.
Compared to the inclusive $\phi \rightarrow \tt$ channel~\cite{TEVtautau},
for the same integrated luminosity the $b\phi \rightarrow b\tt$ channel offers increased 
sensitivity in the low $M_\phi$ region, as it does not suffer from the large 
$Z\to\tau^+\tau^-$ background.

In summary, we have presented results from a search for $b\phi \rightarrow b\tt$
production, resulting in significant portions of the MSSM parameter space being
excluded in two specific scenarios. This analysis is found to be both competitive
and complementary to other searches in the $b\phi \rightarrow b\bb$ and inclusive
$\phi\rightarrow \tt$ channels, hence contributing to the overall sensitivity
at the Tevatron.

%
We thank the staffs at Fermilab and collaborating institutions, 
and acknowledge support from the 
DOE and NSF (USA);
CEA and CNRS/IN2P3 (France);
FASI, Rosatom and RFBR (Russia);
CNPq, FAPERJ, FAPESP and FUNDUNESP (Brazil);
DAE and DST (India);
Colciencias (Colombia);
CONACyT (Mexico);
KRF and KOSEF (Korea);
CONICET and UBACyT (Argentina);
FOM (The Netherlands);
STFC (United Kingdom);
MSMT and GACR (Czech Republic);
CRC Program, CFI, NSERC and WestGrid Project (Canada);
BMBF and DFG (Germany);
SFI (Ireland);
The Swedish Research Council (Sweden);
CAS and CNSF (China);
and the
Alexander von Humboldt Foundation (Germany).


\begin{thebibliography}{99}

%
\bibitem[a]{alton}
Visitor from Augustana College, Sioux Falls, SD, USA.
\bibitem[b]{askew,gershtein}
Visitor from Rutgers University, Piscataway, NJ, USA.
\bibitem[c]{burdin}
Visitor from The University of Liverpool, Liverpool, UK.
\bibitem[d]{hensel,meyer,park,quadt}
Visitor from II. Physikalisches Institut, Georg-August-University,
  G{\"o}ttingen, Germany.
\bibitem[e]{luna-garcia}
Visitor from Centro de Investigacion en Computacion - IPN,
  Mexico City, Mexico.
\bibitem[f]{podesta-lerma}
Visitor from ECFM, Universidad Autonoma de Sinaloa, Culiac\'an, Mexico.
\bibitem[g]{voutilainen}
Visitor from Helsinki Institute of Physics, Helsinki, Finland.
\bibitem[h]{weber}
Visitor from Universit{\"a}t Bern, Bern, Switzerland.
\bibitem[i]{wenger}
Visitor from Universit{\"a}t Z{\"u}rich, Z{\"u}rich, Switzerland.
\bibitem[\ddag]{deceased}
Deceased.

%
\vskip 0.25cm


\bibitem{CHWW}
M.~Carena, S.~Heinemeyer, C.E.M.~Wagner and G.~Weiglein,
Eur. Phys. J. C {\bf 45}, 797 (2006).

\bibitem{bbb}
D0 Collaboration, V.M.~Abazov {\sl et al.}, arXiv:0805.3556 [hep-ex] (2008),
accepted by Phys. Rev. Lett.

\bibitem{TEVtautau}
CDF Collaboration, A.~Abulencia {\sl et al.}, Phys. Rev. Lett. {\bf 96}, 011802 (2006);
D0 Collaboration, V.M.~Abazov {\sl et al.}, Phys. Rev. Lett. {\bf 101}, 071804 (2008).

\bibitem{D0detector}
D0 Collaboration, V.M.~Abazov {\sl et al.}, Nucl. Instrum. Methods Phys. Res. A {\bf 565},
463 (2006).

\bibitem{NewLumi}
T.~Andeen {\sl et al.}, FERMILAB-TM-2365 (2007).

\bibitem{pythia}
T.~Sj\"ostrand {\sl et al.}, Comput. Phys. Commun. {\bf 135}, 238 (2001).

\bibitem{alpgen}
M.~Mangano {\sl et al.}, JHEP {\bf 0307}, 1 (2003).

\bibitem{geant}
R.~Brun and F.~Carminati, CERN Program Library Long Writeup W5013, 1993 (unpublished).

\bibitem{Z-tautau}
D0 Collaboration, V.M.~Abazov {\sl et al.}, Phys. Rev. D {\bf 71}, 072004 (2005) 
[Erratum-ibid. {\bf 77}, 039901(2008)]

\bibitem{jetalg}
G.C.~Blazey {\sl et al.}, arXiv:hep-ex/0005012 (2000).

\bibitem{JLIP}
S.~Greder, FERMILAB-THESIS-2004-28.

\bibitem{Junk}
T.~Junk, Nucl. Instrum. Methods Phys. Res. A {\bf 434}, 435 (1999).

\bibitem{mcfm}
J.~Campbell, R.K.~Ellis, F.~Maltoni and S.~Willenbrock,
Phys. Rev. D {\bf 67}, 095002 (2003).


\bibitem{feynhiggs}
S.~Heinemeyer, W.~Hollik and G.~Weiglein,
Comput. Phys. Commun. {\bf 124}, 76 (2000).

\end{thebibliography}
\end{document}